\documentclass[12pt]{article}
\font\twozero=cmr10 at 20pt
\font\oneeight=cmr10 at 18pt

\newcommand{\vT}{\vphantom{\mbox{\twozero I}}}
\newcommand{\vTm}{\vphantom{\mbox{\oneeight I}}}

\begin{document}

\baselineskip=20pt

\newfont{\elevenmib}{cmmib10 scaled\magstep1}
\newcommand{\preprint}{
    \begin{flushleft}
      \elevenmib Yukawa\, Institute\, Kyoto\\
    \end{flushleft}\vspace{-1.3cm}
    \begin{flushright}\normalsize  \sf
      YITP-05-60\\
      October 2005
    \end{flushright}}
\newcommand{\Title}[1]{{\baselineskip=26pt
    \begin{center} \Large \bf #1 \\ \ \\ \end{center}}}
\newcommand{\Author}{\begin{center}
    \large \bf   R.~Sasaki${}^a$
and K.~Takasaki${}^b$\end{center}}
\newcommand{\Address}
{\begin{center}
  ${}^a$ Yukawa Institute for Theoretical Physics,\\
      Kyoto University, Kyoto 606-8502, Japan\\
      $^b$ Graduate School of  Human and Environmental Sciences \\
   Kyoto University, Kyoto 606-8502, Japan
    \end{center}}
\newcommand{\Accepted}[1]{\begin{center}
    {\large \sf #1}\\ \vspace{1mm}
{\small \sf Accepted for Publication}
    \end{center}}

\preprint
\thispagestyle{empty}
\bigskip\bigskip\bigskip

\Title{Explicit solutions of the classical
Calogero \& Sutherland
systems for any root system}
\Author

\Address

\begin{abstract}
Explicit solutions of the classical Calogero (rational
with/without harmonic confining potential) and Sutherland
(trigonometric potential) systems is obtained by diagonalisation
of certain matrices of simple time evolution.
The method works for Calogero \& Sutherland systems based on any
root system.  It generalises the well-known results by
Olshanetsky and Perelomov for the $A$ type root systems.
Explicit solutions of the (rational and trigonometric) higher Hamiltonian
flows of the integrable hierarchy can be readily obtained in a similar way
for those based on the classical root systems.
\end{abstract}

\section{Introduction}
\label{intro}
\setcounter{equation}{0}

The classical and quantum integrability/solvability of
Calogero-Moser systems \cite{Cal1}--\cite{Cal2} manifests itself
in many guises; the existence of Lax pairs and/or Dunkl operators,
algebraic linearisation, quadratic algebras,
associated integrable spin chains,
`quantised' classical spectra, etc.
Among them, a most intuitive understanding of solvability/integrability
is provided by the fact that explicit solutions of the classical
equations of motion are obtained by diagonalisation of certain matrices
having trivial time evolution as shown by Olshanetsky and Perelomov
\cite{OP3} for the rational  and trigonometric
  potential cases \cite{Suris}.
Their results are for the systems based on the $A$ type roots.
Here we will show that the same results hold universally for
systems based on any root system.
To be more precise, for the rational potential  (without/with harmonic
confining potential) cases, the diagonalisation method works
for any root system including the non-crystallographic ones.
For the trigonometric and hyperbolic potential cases,
we show that the explicit diagonalisation method holds for
any crystallographic root system based on the {\em universal} Lax pair
\cite{bcs2}.
A simpler form of explicit diagonalisation is provided by
the  {\em minimal\/} Lax pair \cite{bcs1,bcs2}, which exists only for
those based on the
$A$, $D$, $E_6$ and $E_7$ root systems. The basic idea of
the explicit solution method is very closely related to the
notion of algebraic linearisation  proved universally for any
root system by  Caseiro-Fran\c{c}iose-Sasaki
\cite{cfs1}. We will follow the notation of \cite{cfs1}
throughout this paper and eq.(a.b) of this paper will be cites
as (Ia.b).
Explicit solutions in terms of diagonalisation
is readily obtained for the {\em higher\/}
(rational and trigonometric)  {\em Hamiltonian flows\/}
belonging to the integrable hierarchy.
This works, however, only for those based on the classical
root systems, $A$, $B$, $C$ and $D$.
The conventional Lax pair in terms of the set of
vector weights ($A$, $C$ and $D$) or
the set of short roots ($B$) is indispensable.

This paper is organised as follows. In section two,
the historical background and the
logical structure of the Calogero-Moser systems
necessary for the present paper are
briefly reviewed. The Hamiltonian and the
universal Lax pair with rational potential
are introduced. The explicit integration
in terms of diagonalisation is achieved by
relating the Lax pair matrices $L$ and $M$
to a matrix $W$ of the same size with
trivial time evolution, $\ddot{W}=0$.
Section three is devoted to the explicit solution of the  systems with
rational plus the  harmonic confining potential.
In section four, we show the explicit solutions of the Sutherland systems,
which have trigonometric/hyperbolic potentials.
In \S4.1 a simple form of explicit diagonalisation
is obtained by re-interpreting the
formulas of the algebraic linearisation method
developed by Caseiro-Fran\c{c}oise-Sasaki
\cite{cfs1}. This is based on the minimal
Lax pair, which exists for those based on $A$, $D$, $E_6$ and $E_7$
root systems. A general treatment of explicit integration of the
Sutherland systems in terms  of the universal Lax pair
is provided in \S4.2. This applies to any crystallographic root system.
Sections five and six are devoted to the problem of explicit
integration of the higher Hamiltonian flows
of the integrable hierarchy.
The rational potential case is discussed in section five and the
trigonometric case in section six.
The final section is for summary and comments.

\section{Rational Potential}
\setcounter{equation}{0}

The integrability of the Calogero-Moser systems has a long history.
Firstly, various types of the integrable potentials are recognised;
starting from the Calogero model \cite{Cal1} with rational ($1/q^2$)
plus a harmonic confining ($q^2$) potential followed by the Sutherland
model \cite{Sut} with a trigonometric ($1/\sin^2q$) potential.
Then the pure rational potential ($1/q^2$) \cite{Mos1} and the
hyperbolic ($1/\sinh^2q$)  \cite{CMR} and the elliptic ($\wp(q)$)
potentials \cite{Cal2} are added to the list of
the integrable potentials.
As seen in many other subjects in mathematical physics,
the quantum groups, the integrable spin chains,
Yang-Baxter equations, etc.,
the elliptic case is the generic one giving all the rest
in various degenerations. However, each degenerate case, the rational,
trigonometric and the hyperbolic,
has its own special properties and merits not
shared by the more generic ones; for example, the algebraic linearisation
\cite{cfs1} of the degenerate Calogero-Moser
systems and the quadratic algebras
\cite{cfs2} for the quantum systems
with the super-integrable rational ($1/q^2$)
potential.
The present article deals with one of such properties of the degenerate
Calogero-Moser systems and it is in fact very closely related to the
algebraic linearisation \cite{cfs1}.
Secondly, the nature of the multi-particle
interactions of the Calogero-Moser
systems is recognised to be governed
by the root systems associated with finite
reflection (Coxeter/Weyl) groups \cite{OP1,OP2}.
The original models \cite{Cal1}--\cite{Cal2}
are all based on the $A$ type root
system related to the symmetric group
$S_N$ with $N$ being the number of the
particles. The $S_N$ is also the Weyl
group of the special unitary group $SU(N)$.
The integrability (the Lax pair) of the
systems based on the classical root
systems ($A$, $B$, $C$ and $D$) is noticed immediately
by Olshanetsky and Perelomov \cite{OP1,OP2},
but the actual demonstration of the integrability of
the Calogero-Moser systems
based on the exceptional \cite{bcs1,DHPH}
and non-crystallographic root systems
\cite{bcs2} took longer years.
The classical universal Lax pair applicable for all types of potentials
and for any root system \cite{bcs2} and the quantum universal Lax pair
applicable for all degenerate
potentials and for any root system \cite{bms}
have been known for some years.

Let us denote by $\Delta$ a root system of rank $r$.
It is a finite set of ${\bf R}^{r}$ vectors
which is invariant under reflections
in the hyperplane perpendicular to each vector in $\Delta$.
A reflection $s_\rho$ in terms of a root $\rho$ is defined by
\begin{equation}
s_\rho(x)=x-\rho\, (\rho^\vee\cdot x),\quad x\in {\bf
R}^r,
\end{equation}
in which $\rho^\vee={2\rho/{\rho^2}}$.
Thus $\Delta$ is characterised by
\begin{equation}
s_\rho(\eta)\in\Delta,\quad \forall \rho,\eta\in\Delta.
\end{equation}

The dynamical variables are  the coordinates $q_{i}\in {\bf
R}$,
$i=1,...,r$ and their canonically conjugate momenta $p_{i}\in {\bf
R}$,
$i=1,...,r$, except for the $A_r$ case in the ordinary embedding,
in which the number of particles is $r+1$.
The Hamiltonian for the classical
Calogero-Moser system with the rational
potential but without the harmonic confining potential  is:

\begin{equation}
     {\cal H}={1\over2}p^2+
        {1\over2}\sum_{\rho\in\Delta_+}
           {g_{|\rho|}^{2}|\rho|^{2}\over{(\rho\cdot q)^2}},
           \label{ratHam}
\end{equation}
in which the real and positive coupling constants $g_{|\rho|}$
  are defined on orbits of the corresponding
Coxeter group.
That is, for the simple Lie algebra cases
$g_{|\rho|}=g$ for all roots in simply-laced models
and  $g_{|\rho|}=g_L$
for long roots and $g_{|\rho|}=g_S$ for
short roots in non-simply laced models.
In order to define Lax pair matrices $L$ and $M$,
let us choose a set of
\(\mathbf{R}^{r}\) vectors
\(\mathcal{R}=\{\alpha, \beta, \ldots,\}\),
$\#\mathcal{R}=\mathcal{D}$,
  permuting under the action of the reflection group:
\begin{equation}
s_\rho(\alpha)\in\mathcal{R},\quad \forall\alpha\in\mathcal{R},\quad
\forall\rho\in\Delta.
\end{equation}
We demand that it consists of
a single orbit of the Coxeter group, for irreducibility.
Then we define  $\mathcal{D}{\times}\mathcal{D}$
matrices indexed by the elements of
$\mathcal{R}$:
\begin{eqnarray}
  p\cdot\hat{H}:& & (p\cdot\hat{H})_{{\alpha}{\beta}} =
    (p\cdot\alpha)\thinspace{\delta}_{{\alpha}{\beta}},
\label{defpH}\\
\hat{s}_{\rho}:& &(\hat{s}_{\rho})_{{\alpha}{\beta}} =
    {\delta}_{\alpha,s_{\rho}({\beta})}. \label{defshat}
\end{eqnarray}

Introduce next the $\mathcal{D}\times \mathcal{D}$
matrices $X$, $L$ and $M$
\cite{bcs1,bcs2}:
\begin{eqnarray}
     X&=&{i}\sum_{\rho\in\Delta_{+}}g_{|\rho|}
        (\rho\cdot\hat{H})\thinspace{1\over{(\rho\cdot
        q)}}\thinspace\hat{s}_{\rho}, \label{Xdef}\\
     L&=& p\cdot\hat{H}+X, \label{Ldef}\\
M&=&-{i\over2}\sum_{\rho\in\Delta_{+}}g_{|\rho|}\thinspace
    {|\rho|^2\over{(\rho\cdot q)^{2}}}\thinspace\hat{s}_{\rho},
\label{Mdef}
    \end{eqnarray}
and a {\em diagonal matrix\/}:
\begin{equation}
     Q=q\cdot\hat{H}:\quad (Q)_{{\alpha}{\beta}} =
     (q\cdot\alpha)\thinspace{\delta}_{{\alpha}{\beta}}.
\label{Qdef}
\end{equation}
Here  $L$ and $Q$ are hermitian $L^\dagger=L$, $Q^\dagger=Q$ and
$M$ is anti-hermitian
$M^\dagger=-M$.

As shown in \cite{cfs1}  (I2.7a), (I2.7b)
the time evolution of the matrix $L$
along the flow of the Hamiltonian
(\ref{ratHam}) displays the following equations:
\begin{eqnarray}
{\partial{L}\over {\partial t}}&=& [L,M], \label{Laxpair}\\
{\partial Q\over{\partial t}}&=&[Q,M] + L. \label{Qevol}
\end{eqnarray}

Next let us define another $\mathcal{D}\times \mathcal{D}$
unitary matrix
$U(t)$ by the linear equation and the initial condition:
\begin{equation}
{\partial{U}\over{\partial t}}=UM,\qquad U(0)=1_\mathcal{D},
\label{Udef}
\end{equation}
in which $1_\mathcal{D}$ is the
$\mathcal{D}\times \mathcal{D}$ unit matrix.
The final step is the introduction of $W(t)$:
\begin{equation}
W(t)\equiv U(t)Q(t)U^{-1}(t),
\label{Wdef}
\end{equation}
which has a simple time evolution
\begin{eqnarray}
\dot{W}&=& U(\dot{Q}-[Q,M])U^{-1}=ULU^{-1},
\label{dotW}\\
\ddot{W}&=& U(\dot{L}-[L,M])U^{-1}=0.
\label{ddotW}
\end{eqnarray}
The solution is
\begin{equation}
W(t)=W(0)+t\,\dot{W}(0),
\label{Wsol}
\end{equation}
with the initial values
\begin{equation}
W(0)=Q(0),\quad \dot{W}(0)=L(0),
\end{equation}
which are determined by the initial values of the
canonical variables $q_j(0)$, $p_j(0)$, $j=1,\ldots, r$.
Due to the defining relation of $W(t)$ in terms of
the {\em diagonal matrix\/} $Q(t)$
(\ref{Wdef}),  the solution $\{q(t)\}$ of the  the canonical
equations of motion
\begin{equation}
{\partial {q}_j\over{\partial t}}
={\partial\mathcal{H}\over{\partial p_j}},
\qquad
{\partial {p}_j\over{\partial t}}
=-{\partial\mathcal{H}\over{\partial q_j}},
\qquad j=1,\ldots,r,
\end{equation}
with the above Hamiltonian (\ref{ratHam}),
is simply obtained
by {\em diagonalising\/} the above matrix solution (\ref{Wsol}).
The conjugate momenta $\{p(t)\}$ are obtained by differentiation
$p_j(t)={\partial{q}_j(t)/{\partial t}}$.

As promised, this is the universal proof applicable for any root
system including the non-crystallographic one.
The spectrum of $W(t)$ (\ref{Wsol}) is highly constrained,
since its dimension $\mathcal{D}$ is usually much greater than the
degree of freedom $r$.
The high symmetry of the spectrum is guaranteed by the
Coxeter invariance of the theory:
\begin{eqnarray}
\mathcal{H}(s_\rho(p),s_\rho(q))&=&\mathcal{H}(p,q),\qquad
\forall\rho\in\Delta,
\label{coxinvHam}\\
L(s_{\rho}(p),s_{\rho}(q))&=&\hat{s}_\rho
L(p,q)\hat{s}_\rho,\quad M(s_{\rho}(q))=\hat{s}_\rho
M(q)\hat{s}_\rho.
\label{lmcoxcov}
\end{eqnarray}
The original proof of the explicit integration of the
$A$ type systems by Olshanetsky and Perelomov \cite{OP3}
is the very special case in which the spectrum of $W(t)$
(\ref{Wsol}) is not constrained.
Our proof reduces to that of \cite{OP3} when $\Delta=A_r$ and
the set of vector weights is chosen as $\mathcal{R}={\bf V}$,
$\#\mathcal{R}=\mathcal{D}=r+1$.

\section{Rational  with Harmonic Confining Potential}
\label{period}
\setcounter{equation}{0}
        The Hamiltonian is now:
\begin{equation}
     {\cal H}_{\omega}={1\over2}p^2+{1\over2}\omega^2q^2+
        {1\over2}\sum_{\rho\in\Delta_+}
           {g_{|\rho|}^{2} |\rho|^{2}\over{(\rho\cdot
           q)^2}}.
           \label{calham}
\end{equation}
With the same matrices introduced above
in the preceding section,
the time evolution displays (I3.2a), (I3.2b):
\begin{eqnarray}
{\dot{L}}&=& [L,M]-{\omega}^{2}Q, \label{Ldotomega}\\
\dot{Q}&=&[Q,M] + L. \label{Qml}
\end{eqnarray}

With the same definition of the unitary matrix $U(t)$ as above
(\ref{Udef}),
  the matrix
\begin{equation}
W(t)\equiv U(t)Q(t)U^{-1}(t),
\label{Wdef2}
\end{equation}
evolves harmonically in time:
\begin{eqnarray}
\dot{W}&=& U(\dot{Q}-[Q,M])U^{-1}=ULU^{-1}\\
\ddot{W}&=& U(\dot{L}-[L,M])U^{-1}=-\omega^2 W.
\end{eqnarray}
The solution is
\begin{equation}
W(t)=\cos\omega t\,W(0)+\omega^{-1}\sin\omega t\,\,\dot{W}(0),
\label{Wsol2}
\end{equation}
with the initial values
\begin{equation}
W(0)=Q(0),\quad \dot{W}(0)=L(0).
\end{equation}
Again the explicit solution $\{q(t)\}$ is obtained by
{\em diagonalising\/} the above matrix $W(t)$ (\ref{Wsol2})
with the harmonic time dependence.

\section{Trigonometric Potential}
\label{trig}
\setcounter{equation}{0}

         The Hamiltonian of the trigonometric (Sutherland)
model \cite{Sut}  writes:
\begin{equation}
     {\cal H}={1\over2}p^2+
        {1\over2}\sum_{\rho\in\Delta_+}
           {g_{|\rho|}^{2}|\rho|^{2}\over{\sin^{2}(\rho\cdot
q)}}.
           \label{SutHam}
\end{equation}
         In order to get the hyperbolic case
it suffices to change  $\sin$
into $\sinh$. In the following, we only demonstrate the
explicit
integration of the trigonometric case. The hyperbolic case can
be deduced easily by the above replacement.

Two types of Lax pairs are known \cite{bcs1,bcs2}
  for the trigonometric cases:
the minimal and the universal Lax pairs.
While the latter, the universal lax pair,
applies to any crystallographic root system,
the former, the minimal Lax pair, requires $\mathcal{R}$
to be the set of minimal weights,
which exists only for the $A$, $D$,
$E_6$ and $E_7$ root systems.
Let us start with the minimal Lax pair which has
simpler structure thanks to the restriction to
the minimal weights, satisfying the condition:
\begin{equation}
\mu:\ \mbox{minimal weight}\ \Leftrightarrow \
\alpha^\vee\cdot \mu=0,\pm1,\quad
\forall \alpha\in\Delta.
\end{equation}

\subsection{Minimal Lax pair}
\label{min}

         We consider the matrices \cite{bcs1,bcs2}:
\begin{eqnarray}
L&=&p\cdot\hat{H}+X, \label{Ldef3}\\
X&=&{ i}\sum_{\rho\in\Delta_{+}}g_{|\rho|}\thinspace
    (\rho\cdot\hat{H})\thinspace{1\over{\sin(\rho\cdot
    q)}}\thinspace\hat{s}_{\rho}, \label{trigX}\\
M&=&-{ i\over2}\sum_{\rho\in\Delta_{+}}g_{|\rho|}\thinspace
    {|\rho|^2\cos(\rho\cdot q)\over{\sin^{2}(\rho\cdot q)}}
    \left(\hat{s}_{\rho}-1_\mathcal{D}\right)+
    i\sum_{\rho\in\Delta_{+}}g_{|\rho|}
{(\rho\cdot \hat{H})^2\over{\sin^2(\rho\cdot q)}},
\label{trigM}
\end{eqnarray}
and {\em diagonal matrices\/}:
\begin{eqnarray}
Q&=&q\cdot\hat{H}:\quad (Q)_{{\alpha}{\beta}} =
(q\cdot\alpha)\thinspace{\delta}_{{\alpha}{\beta}},
  \label{(5.2d)}\\
R&=&{{\rm e}}^{2{ i}Q}. \label{(5.2e)}
\end{eqnarray}
Again $L$ and $Q$ are hermitian $L^\dagger=L$, $Q^\dagger=Q$ and
$M$ is anti-hermitian
$M^\dagger=-M$. Thus $R$ is unitary.

\bigskip
As shown in \cite{cfs1} (I5.3a), (I5.3b),  when the
root system admits a minimal representation,
and $\mathcal{R}$ being the set of minimal weights, the time
evolution along the flow of the Hamiltonian (\ref{SutHam})
displays:
\begin{eqnarray}
{\partial {L}\over{\partial t}}&=&[L,M], \label{trigLax}\\
{\partial {R}\over{\partial t}}&=&[R,M]+
{i}\left(RL+LR\right).
\label{TrigRLM}
\end{eqnarray}

With the same definition of the unitary matrix $U(t)$ as above
(\ref{Udef}),
\begin{equation}
{\partial{U}\over {\partial t}}=UM,\qquad U(0)=1_\mathcal{D},
\label{UdefTrig}
\end{equation}
we introduce a matrix
\begin{equation}
\mathcal{W}(t)=U(t)R(t)\,U(t)^{-1}=U(t)\,{{\rm e}}^{2{
i}Q(t)}\,U(t)^{-1}.
\end{equation}
It satisfies a simple first order linear differential equation
\begin{eqnarray}
{\partial{\mathcal{W}}\over {\partial t}}&=
&U(\partial{R}/\partial t-[R,M])U^{-1}
=i\,U(RL+LR)U^{-1}\\
&=&i\left(\mathcal{W}\,UL\,U^{-1}+UL\,U^{-1}\mathcal{W}\right),
\end{eqnarray}
since as in (\ref{dotW}), (\ref{ddotW}), $UL\,U^{-1}$ is a
constant matrix:
\begin{eqnarray}
{\partial \over{\partial t}}\left(UL\,U^{-1}\right)&=&
U(\partial{L}/\partial t-[L,M])U^{-1}=0,\\ U(t)L(t)U(t)^{-1}&=&L(0).
\end{eqnarray}
The solution is
\begin{equation}
\mathcal{W}(t)={\rm e}^{itL(0)}{\rm e}^{2iQ(0)}{\rm e}^{itL(0)}.
\label{Sutsol}
\end{equation}
By {\em diagonalising\/} the above matrix solution,
we obtain the explicit solution $\{q(t)\}$ of the
classical Sutherland model (\ref{SutHam}).
One might naturally wonder if the coordinates $\{q(t)\}$ could
be determined uniquely from the unitary matrix (\ref{Sutsol}).
The answer is affirmative since the motion is always restricted
to one of the Weyl alcoves due to the periodicity and singularity
of the potential.
Near the boundary of a Weyl alcove, for example at $\rho\cdot
q=0$, $\rho\in\Delta$, the singularity of the potential
$\sim 1/(\rho\cdot q)^2$ can never be surpassed classically.
Therefore if $\{q(0)\}$ is in the principal Weyl alcove,
\begin{equation}
PW_T=\{q\in{\bf R}^r|\rho\cdot q>0,\quad \rho\in\Pi,\quad
\rho_h\cdot q<\pi\},
\end{equation}
$\{q(t)\}$ will always remain there. Here $\Pi$ is the set of
the simple roots and $\rho_h$ is the highest weight.
This  removes any ambiguity in determining
$\{q(t)\}$ from the eigenvalues of (\ref{Sutsol}).
As in the rational potential cases, the spectrum of
$\mathcal{W}(t)$ is highly constrained as a consequence of the
Weyl invariance (\ref{coxinvHam}), (\ref{lmcoxcov}).

\subsection{Universal Lax pair}
\label{uni}
The universal Lax pair has $\cot(\rho\cdot q)$ function in
$L$ instead of $1/\sin(\rho\cdot q)$
in (\ref{trigX})
\begin{eqnarray}
L&=&p\cdot\hat{H}+X, \label{Ldef4}\\
X&=&{ i}\sum_{\rho\in\Delta_{+}}g_{|\rho|}\thinspace
    (\rho\cdot\hat{H})\thinspace{\cot(\rho\cdot
    q)}\thinspace\hat{s}_{\rho}, \label{trigX2}\\
M&=&-{ i\over2}\sum_{\rho\in\Delta_{+}}g_{|\rho|}\thinspace
    {|\rho|^2\over{\sin^{2}(\rho\cdot q)}}
    \left(\hat{s}_{\rho}-1_D\right),
\label{trigMuni}
\end{eqnarray}
which satisfy $\partial{L}/\partial t=[L,M]$
for the Hamiltonian flow
but the additional equation (\ref{TrigRLM})
takes a  different form.

For $\mathcal{R}$ being the set of minimal weights,
  it reads
\begin{equation}
{\partial{R}\over{\partial t}}=[R,M]+{i}\left(\vTm
R(L+K)+(L-K)R\right),
  \label{TrigRLMuni1}
\end{equation}
in which $K$ is a non-negative constant
matrix  commuting with $M$:
\begin{equation}
K\equiv
\sum_{\rho\in\Delta_+}g_{|\rho|}(\rho\cdot\hat{H})
(\rho^\vee\!\cdot\hat{H})
\hat{s}_\rho,\quad [K,M]=0.
\label{Kexp}
\end{equation}
It is a very important quantity in Calogero-Moser
systems appearing in many contexts.
For example, it is a commutator of  $Q$ (\ref{Qdef})
and the rational Lax matrix
$L$  (\ref{Ldef}),
(\ref{Xdef}) (see (4.36) of \cite{bms} and (2.40) of \cite{cs}):
\begin{equation}
[Q,L]=iK.
\label{Kdef}
\end{equation}
It should be noted that if $K$ is defined as above,
the expression (\ref{Kexp}) is {\em universal\/},
that is valid for any root system $\Delta$ and
any choice of $\mathcal{R}$.
Various properties of the $K$ matrix,
whose eigenvalues are all `integers',
are discussed in detail by Corrigan-Sasaki,
in the Appendix of \cite{cs}.

For $\mathcal{R}$ being the set of all roots $\Delta$
(for the simply-laced root systems)
or the set of short  roots $\Delta_S$
(for non simply-laced root systems)  and also the
set of vector weights (${\bf V}$) for the $C$,
the relation corresponding to (\ref{TrigRLM})
and (\ref{TrigRLMuni1}) reads
\begin{equation}
{\partial{R}\over{\partial t}}=[R,M]+{i}\left(\vTm
R(L+\tilde{K})+(L-\tilde{K})R\right),
\label{TrigRLMuni2}
\end{equation}
in which $\tilde{K}$ is another constant matrix
commuting with $M$,
\begin{equation}
\tilde{K}=
\sum_{\rho\in\Delta_{+}}g_{|\rho|}|\rho\cdot\hat{H}|
     \hat{s}_{\rho},\qquad [\tilde{K},{M}]=0,
     \label{Ktilde}
\end{equation}
introduced by Corrigan-Sasaki, as (5.32) of \cite{cs}.

Now the explicit solution of the
Sutherland system is achieved for any
crystallographic root system,
since one can choose at least
one Lax pair satisfying (\ref{TrigRLMuni2}).
We proceed as before by
defining the unitary matrix $U(t)$ by
(\ref{UdefTrig}) and introduce
a  matrix
\begin{equation}
\mathcal{W}(t)=U(t)R(t)\,U(t)^{-1}=U(t)\,{{\rm e}}^{2{
i}Q(t)}\,U(t)^{-1}.
\end{equation}
It satisfies a simple first order linear differential equation
\begin{eqnarray}
{\partial{\mathcal{W}}\over{\partial t}}&=&
U(\partial{R}/\partial t-[R,M])U^{-1}
=i\,U(\vT R(L+\tilde{K})+(L-\tilde{K})R)U^{-1}\\
&=&i\left(\mathcal{W}\,U(L+\tilde{K})\,U^{-1}
+U(L-\tilde{K})\,U^{-1}\mathcal{W}\right),
\end{eqnarray}
since as in (\ref{dotW}), (\ref{ddotW}),
$U(L\pm\tilde{K})\,U^{-1}$ is a
constant matrix:
\begin{eqnarray}
{d\over{dt}}\left(U(L\pm\tilde{K})\,U^{-1}\right)&=&
U(\partial{L}/\partial t-[L,M])U^{-1}=0,\\
U(t)(L(t)\pm\tilde{K})U(t)^{-1}&=&L(0)\pm \tilde{K}.
\end{eqnarray}
The solution is
\begin{equation}
\mathcal{W}(t)=
{\rm e}^{it(L(0)-\tilde{K})}{\rm e}^{2iQ(0)}
{\rm e}^{it(L(0)+\tilde{K})}.
\label{Sutsoluni}
\end{equation}
By {\em diagonalising\/} the above matrix solution,
we obtain the explicit solution $\{q(t)\}$ of the
classical Sutherland system (\ref{SutHam}) for any root system.

The very fact that $\tilde{K}$ ($K$) commutes with
$M$ simply means that a spectral parameter
$\lambda$ can be introduced trivially into the Lax pair for
degenerate potentials \cite{bcs2}:
\begin{equation}
L_\lambda\equiv L+\lambda\tilde{K},
\quad \dot{L}_\lambda=[L_\lambda,M].
\end{equation}

\section{Rational Higher  Flows }
\setcounter{equation}{0}

The integrable hierarchy of the
Calogero-Sutherland systems consists of
Hamiltonians generated by higher conserved
quantities, which are constructed,
for example, from the trace
of the higher powers of the $L$ matrix,
$\mathcal{H}_n\propto\mbox{Tr}(L^{2n})$.
The method of explicit integration
as described in the preceding sections
applies also to these higher Hamiltonian flows,
as shown by Suris \cite{Suris} for the $A$ type root systems with
the conventional Lax pair, $\mathcal{R}={\bf V}$.
However, in contrast to the basic Calogero-Sutherland flows, it works
only for those systems based on the classical root systems, the $A$,
$B$, $C$ and $D$. Let us denote by $\mathcal{R}$ the set of vector
weights ${\bf V}$, for the
$A$, $C$ and $D$ root systems and the set of short roots
$\Delta_S$ for the
$B$ root system.  These particular sets $\mathcal{R}$ have unique
orthogonality property
\begin{equation}
\mbox{if}\ \mu\neq\pm\nu,\quad \mu\cdot\nu=0,\quad
\forall\mu,\nu\in\mathcal{R},
\end{equation}
which endows a very special structure to the Lax pair
represented on $\mathcal{R}$.
The dimensions of the corresponding Lax matrices are
$\mathcal{D}=r+1$ for the $A_r$ and $\mathcal{D}=2r$
for the $B_r$, $C_r$ and $D_r$.
It is through these special Lax matrices that the
explicit integration of the higher
rational and trigonometric flows is realised.

Let us start with the explicit forms of the rational $L$ matrices:
\begin{eqnarray}
(A):&& \qquad\qquad L,\hspace*{18mm} L_{jk}=p_j\delta_{jk}
+ig(1-\delta_{jk})/(q_j-q_k),\\
(B):&&
L=\pmatrix{A &B\cr
-B&-A},\quad A_{jk}=p_j\delta_{jk}
+ig_L(1-\delta_{jk})/(q_j-q_k),\nonumber\\
&& \phantom{L=\pmatrix{A &B\cr
-B&-A},} \quad  B_{jk}=i(g_S/q_j)\delta_{jk}
+ig_L(1-\delta_{jk})/(q_j+q_k).
\end{eqnarray}
The rational $C$ system will not be discussed since
it is equivalent to the rational $B$ system.
The rational $D$ system is obtained by
constraining $g_S=0$ in the rational $B$ system.

The higher Hamiltonians are
\begin{eqnarray}
(A): &&\mathcal{H}_n=\mbox{Tr}(L^{n+1})/(n+1),\qquad n\ge1,
\label{nAratham}\\
(B,D):&& \mathcal{H}_n=\mbox{Tr}(L^{2n})/(4n),
\qquad\qquad n\ge1.
\label{nBratham}
\end{eqnarray}
The lowest $\mathcal{H}_1$  is the original
Hamiltonian (\ref{ratHam}).
The basic idea is to rewrite the Hamiltonian flow
\begin{equation}
{\partial {q}_j\over{\partial t_n}}=
{\partial\mathcal{H}_n\over{\partial p_j}},
\qquad
{\partial {p}_j\over{\partial t_n}}=
-{\partial\mathcal{H}_n\over{\partial q_j}},
\label{nthflow}
\end{equation}
into equivalent matrix forms
\begin{eqnarray}
{\partial L\over{\partial t_n}}&=& [L,M_n],
\label{Laxpairnth}\\
{\partial Q\over{\partial t_n}}&=&[Q,M_n] + L^{n}\
\left(L^{2n-1}\right),
\label{Qevolnth}
\end{eqnarray}
as in the lowest flow (\ref{Laxpair}), (\ref{Qevol}).

In contrast to the lowest flow case in which
the explicit form of $M$ is given (\ref{Mdef}), we can
interpret part of (\ref{Laxpairnth}) and
(\ref{Qevolnth}) as determining $M_n$.
The diagonal part of the $Q$ equation (\ref{Qevolnth})
is equivalent to the
first half of the canonical equations (\ref{nthflow}).
The off-diagonal part  of the $Q$ equation (\ref{Qevolnth})
determines the off-diagonal part of $M_n$
completely:
\begin{equation}
(M_n)_{\mu\nu}=-{(L^n)_{\mu\nu}/{q\cdot(\mu-\nu)}}\quad
\left(-{(L^{2n-1})_{\mu\nu}/{q\cdot(\mu-\nu)}}\right),
\qquad \mu\neq\nu.
\end{equation}
Whereas the diagonal part of $M_n$ does not enter
the the $Q$ equation (\ref{Qevolnth}),
it can be determined from the off-diagonal
part of the Lax equation.
The result is very simple:
\begin{eqnarray}
(M_n)_{\mu\mu}=-\sum_{\nu\neq \mu}(M_n)_{\nu \mu}=
-\sum_{\nu \neq \mu}(M_n)_{\mu \nu},
\quad M_n^\dagger=-M_n.
\end{eqnarray}
The proof that the diagonal part of the higher flow Lax equation
  (\ref{Laxpairnth})
is  equivalent to the second  half of
the canonical equations (\ref{nthflow})
goes almost parallel to that of the lowest flow.

After the equivalence of the canonical equations (\ref{nthflow})
  with the two matrix equations  (\ref{Laxpairnth})
and (\ref{Qevolnth}) is
established, the explicit integration
by diagonalisation is straightforward. Let
us define a $\mathcal{D}\times \mathcal{D}$ unitary matrix
$U_n(t_n)$ by the linear equation and the initial condition:
\begin{equation}
{\partial{U_n}\over{\partial t_n}}=U_nM_n,
\qquad U_n(0)=1_\mathcal{D}.
\label{Udefnth}
\end{equation}
Then a matrix  function $W_n(t_n)$, defined by
\begin{equation}
W_n(t_n)\equiv U_n(t_n)Q(t_n)U_n^{-1}(t_n),
\label{Wdefnth}
\end{equation}
has a simple time evolution
\begin{eqnarray}
&&{\partial{W_n}\over{\partial t_n}}=
U_n(\partial{Q}/\partial t_n-[Q,M_n])U_n^{-1}
=\left(U_nLU_n^{-1}\right)^n
\quad \left((U_nLU_n^{-1})^{2n-1}\right),
\label{dotWnth}\\
&&{\partial\over{\partial t_n}}\left(U_nL\,U_n^{-1}\right)=
U_n(\partial{L}/\partial t_n-[L,M_n])U_n^{-1}=0,\\
&&\hspace*{28mm} \Longrightarrow U_n(t_n)L(t_n)U_n(t_n)^{-1}=L(0).
\label{ddotWnth}
\end{eqnarray}
The solution is
\begin{equation}
W_n(t_n)=W_n(0)+t_n\,\partial{W_n}(0)/\partial t_n,
\label{Wsolnth}
\end{equation}
with the initial values
\begin{equation}
W_n(0)=Q(0),\quad\partial{W_n}(0)/\partial t_n
=L(0)^n\quad (L(0)^{2n-1}),
\end{equation}
which are determined by the initial values of the
canonical variables $q_j(0)$, $p_j(0)$, $j=1,\ldots, r$.
Due to the defining relation of $W_n(t_n)$ in terms of
the {\em diagonal matrix\/} $Q(t_n)$
(\ref{Wdefnth}),  the solution $\{q(t_n)\}$ of the  the canonical
equations of motion (\ref{nthflow})
with the above Hamiltonian (\ref{nAratham}) or  (\ref{nBratham}),
is simply obtained
by {\em diagonalising\/} the above matrix solution (\ref{Wsolnth}).
Determination of the conjugate momenta  $\{p(t_n)\}$ requires
solution of the second half of the canonical
equations of motion (\ref{nthflow}),
which are now {\em algebraic\/} since
$\{\partial q/\partial t_n\}$ are
now known functions of time.
Extension to the generic higher flows of the hierarchy
\begin{equation}
\mathcal{H}=\sum_{n}c_n\mathcal{H}_n,\quad c_n:\ const.,
\label{genham}
\end{equation}
is straightforward since the matrix equations
  (\ref{Laxpairnth}) and (\ref{Qevolnth}) are linear in $M_n$.
However, some higher flows cannot be treated this way.
For example,
in the $D_r$ ($r$: odd) theory, there exists another conserved
quantity (Hamiltonian) of the form $p_1p_2\cdots p_r+\cdots$,
which cannot be written as (\ref{genham}).

\section{Trigonometric Higher Flows}
\setcounter{equation}{0}

The basic logics of the explicit integration of the trigonometric
higher flows is almost the same as that of the rational higher
flows, except that we have to consider two different types of
Lax pairs; the minimal and the universal.
So we just write down the key formulas without detailed derivation.

\subsection{Minimal Lax Pair}
We discuss the explicit integration of the trigonometric
higher flows of the $A$ and $D$ theory in terms of the minimal
Lax pair, although the formulation in terms of the universal
Lax pair works well for them, too.

The explicit forms of the trigonometric minimal $L$ matrices are:
\begin{eqnarray}
(A):&& \qquad\qquad L,\hspace*{18mm} L_{jk}=
p_j\delta_{jk}+ig(1-\delta_{jk})/\sin(q_j-q_k),\\
(D):&&
L=\pmatrix{A &B\cr
-B&-A},\quad A_{jk}=p_j\delta_{jk}
+ig(1-\delta_{jk})/\sin(q_j-q_k),
\nonumber\\
&& \phantom{L=\pmatrix{A &B\cr
-B&-A},} \quad\  B_{jk}=ig(1-\delta_{jk})/\sin(q_j+q_k).
\end{eqnarray}
The higher Hamiltonians take exactly the
same form as (\ref{nAratham})
and (\ref{nBratham}).
The lowest $\mathcal{H}_1$
is the original Hamiltonian (\ref{SutHam}).
We rewrite the higher Hamiltonian flow (\ref{nthflow})
into equivalent matrix forms
\begin{eqnarray}
{\partial L\over{\partial t_n}}&=& [L,M_n],
\label{minLaxpairnth}\\
{\partial R\over{\partial t_n}}&=&[R,M_n] +{i}(\vT RL^n+L^nR)
\quad \left(\vTm {i}(RL^{2n-1}+L^{2n-1}R)\right) ,
\label{minRevolnth}
\end{eqnarray}
as in the lowest flow (\ref{trigLax}), (\ref{TrigRLM}).
The off-diagonal part of $M_n$ is
\begin{equation}
(M_n)_{\mu\nu}=-{(L^n)_{\mu\nu}\cot[q\cdot(\mu-\nu)]}\quad
\left(-{(L^{2n-1})_{\mu\nu}\cot[q\cdot(\mu-\nu)]}\right),
\qquad \mu\neq\nu.
\end{equation}
The diagonal part is
\begin{equation}
(M_n)_{\mu\mu}=
\sum_{\nu\neq\mu}{(L^n)_{\nu\mu}/\sin[q\cdot(\nu-\mu)]}
=\sum_{\nu\neq\mu}{(L^n)_{\mu\nu}/\sin[q\cdot(\mu-\nu)]},
\quad M_n^\dagger=-M_n.
\end{equation}
The $\mathcal{D}\times \mathcal{D}$ matrix
$\mathcal{W}_n(t_n)$ obeys   simple time evolution:
\begin{eqnarray}
\mathcal{W}_n(t_n)&=&U_n(t)R(t_n)\,U_n(t_n)^{-1}=
U_n(t_n)\,{{\rm e}}^{2{
i}Q(t_n)}\,U_n(t_n)^{-1},\\
&=&
{\rm e}^{it_nL(0)^n}{\rm e}^{2iQ(0)}{\rm e}^{it_nL(0)^n}
\quad \left( {\rm e}^{it_nL(0)^{2n-1}}{\rm e}^{2iQ(0)}{\rm
e}^{it_nL(0)^{2n-1}}\right).
\label{minnexpl}
\end{eqnarray}
By {\em diagonalising\/} the above matrix solution (\ref{minnexpl}),
we obtain the explicit solution $\{q(t_n)\}$ of the
higher flows of the Sutherland system  (\ref{nAratham})
and (\ref{nBratham}) for the $A$ and $D$
  root systems.

\subsection{Universal Lax Pair}
The explicit integration of the higher flows of the $B$ and
$C$ Sutherland systems is achieved in terms of the universal
Lax pairs based on the set of short roots ($\mathcal{R}=\Delta_S$)
for $B$ and the set of vector weights  ($\mathcal{R}=\mathbf{V}$)
for $C$. For the rank $r$ system both have $\mathcal{D}=2r$.

The Lax matrix $L$ and the constant matrix $\tilde{K}$
(\ref{Ktilde}) are:
\begin{eqnarray}
&&\hspace*{20mm}
L=\pmatrix{A &B\cr
-B&-A},\qquad\qquad
\tilde{K}=\pmatrix{S &T\cr
T&S},\\
\hspace*{-10mm}(B):
A_{jk}&=&p_j\delta_{jk}+ig_L(1-\delta_{jk})\cot(q_j-q_k),
\qquad\qquad
S_{jk}=g_L(1-\delta_{jk}),\\
B_{jk}&=&ig_S\cot q_j\,
\delta_{jk}+ ig_L(1-\delta_{jk})\cot(q_j+q_k),
\quad
T_{jk}=g_S\delta_{jk}+ g_L(1-\delta_{jk}),\\
\hspace*{-10mm}(C):
A_{jk}&=&p_j\delta_{jk}+ig_S(1-\delta_{jk})\cot(q_j-q_k),
\qquad\qquad
S_{jk}=g_S(1-\delta_{jk}),\\
B_{jk}&=&2ig_L\cot 2q_j\, \delta_{jk}
+ ig_S(1-\delta_{jk})\cot(q_j+q_k),
\quad
T_{jk}=2g_L\delta_{jk}+ g_S(1-\delta_{jk}).
\end{eqnarray}
It is easy to see
\begin{eqnarray}
{\rm e}^{iQ}(L+\tilde{K}){\rm e}^{-iQ}&=&
{\rm e}^{-iQ}(L-\tilde{K}){\rm e}^{iQ},\\
\Rightarrow \mbox{Tr}(L+\tilde{K})^n&=&\mbox{Tr}(L-\tilde{K})^n,
\end{eqnarray}
which are conserved quantities of the  Sutherland flow
(\ref{SutHam}). It differs from the usual one Tr$(L^{n})$ by
a linear combination of lower order conserved quantities.
The canonical equations of the higher flow Hamiltonian
\begin{equation}
\mathcal{H}_n=\mbox{Tr}\left((L\pm\tilde{K})^{2n}\right)/(4n)
\label{Lkuni}
\end{equation}
are equivalent to the matrix equations
\begin{eqnarray}
{\partial L\over{\partial t_n}}&=& [L,M_n],
\label{minLaxpairnth2}\\
{\partial R\over{\partial t_n}}&=&[R,M_n]
  +{i}\left(\vTm R(L+\tilde{K})^{2n-1}
+(L-\tilde{K})^{2n-1}R\right).
\label{uniRevolnth}
\end{eqnarray}
The off-diagonal part of $M_n$ is
\begin{equation}
(M_n)_{\mu\nu}=-\left[
e^{iq\cdot(\mu-\nu)}{(L+\tilde{K})^n}_{\mu\nu}
+e^{-iq\cdot(\mu-\nu)}{(L-\tilde{K})^n}_{\mu\nu}\right]
/\sin[q\cdot(\mu-\nu)],\quad
  \mu\neq\nu.
\end{equation}
The diagonal part is
\begin{eqnarray}
(M_n)_{\mu\mu}=-\sum_{\nu\neq \mu}(M_n)_{\nu \mu}=
-\sum_{\nu \neq \mu}(M_n)_{\mu \nu},
\quad M_n^\dagger=-M_n.
\end{eqnarray}
The $\mathcal{D}\times \mathcal{D}$ matrix $\mathcal{W}_n(t_n)$ obeys
simple time evolution:
\begin{eqnarray}
\mathcal{W}_n(t_n)&=&U_n(t)R(t_n)\,U_n(t_n)^{-1}=
U_n(t_n)\,{{\rm e}}^{2{
i}Q(t_n)}\,U_n(t_n)^{-1},\\
&=&
  {\rm e}^{it_n(L(0)-\tilde{K})^{2n-1}}{\rm e}^{2iQ(0)}{\rm
e}^{it_n(L(0)+\tilde{K})^{2n-1}}.
\label{uninexpl}
\end{eqnarray}
By {\em diagonalising\/} the above
matrix solution (\ref{uninexpl}),
we obtain the explicit solution $\{q(t_n)\}$ of the
higher flows of the Sutherland system  (\ref{Lkuni})
  for the $B$ and $C$
  root systems.

\section{Summary and Comments}
\setcounter{equation}{0}

Explicit integration of the Calogero and Sutherland systems
by means of diagonalisation is demonstrated for any root system,
the exceptional as well as the classical and the
non-crystallographic.
It is based on the universal Lax pair for the degenerate
potentials, that is the rational with/without the harmonic
confining potential and the trigonometric/hyperbolic potentials.
As emphasised in the text, it is very closely related to the
concept of algebraic linearisation by Caseiro-Fran\c{c}oise-Sasaki
\cite{cfs1}.
The method is extended to the higher Hamiltonian flows
of the rational and trigonometric/hyperbolic interactions.
In contrast to the basic Calogero-Sutherland flows,
the applicability is limited to those systems based on the
classical root systems, the $A$, $B$, $C$ and $D$ root systems.

The theory of explicit integration of higher Hamiltonian flows
is very closely
related to the dynamical $r$-matrix \cite{Suris,avantalon,abt} 
and the Hamiltonian reduction \cite{kks,OP2}.  
In the case of the most classical rational potential of 
the A type, the method of Hamiltonian reduction starts from 
the large phase space of the matrix dynamical variable $W$ 
and its conjugate momentum variable $Z$, which are both 
assumed to be Hermitian.   The Hamiltonians 
\begin{eqnarray}
\mathcal{H}_n = \mbox{Tr}(Z^{n+1})/(n+1) 
\end{eqnarray}
generate the flows 
\begin{eqnarray}
  \frac{\partial W}{\partial t_n} = Z^n,\qquad 
  \frac{\partial Z}{\partial t_n} = 0. 
\end{eqnarray}
This Hamiltonian system in invariant under the action 
$(W,Z) \to (UWU^{-1}, UZU^{-1})$ of unitary matrices $U$.  
The reduced phase space is obtained by imposing 
the constraint 
\begin{eqnarray}
  [W,Z] = iK 
\end{eqnarray}
and factoring out the constrained phase space 
by residual symmetries (i.e., by the group of 
unitary matrices that commute with $K$).  
The $(Q,L)$ pair (\ref{Kdef}) is nothing but a representative 
of a point of the reduced phase space, which is 
connected with the point $(W,Z)$ of the large 
phase space by a ($t$-dependent) unitary matrix 
$U$ as 
\begin{eqnarray}
  Q = U^{-1}WU, \qquad 
  L = U^{-1}ZU. 
\end{eqnarray}
The linear flows of $(W, Z)$ are thereby mapped to 
the Calogero flows of $(Q,L)$.  This is the way 
 to understand the rational Calogero system of 
the A type as a Hamiltonian reduction \cite{kks}; 
a similar interpretation has been proposed for 
a few other cases \cite{OP2}.  The dynamical $r$-matrix 
has been constructed in this framework of Hamiltonian 
reduction \cite{Suris}.  We expect that all the cases 
discussed in this paper can be treated in the same way.

\section*{Acknowledgements}
We thank F. Calogero and J.-P. Fran\c{c}oise
for illuminating discussion.
We are supported in part by Grant-in-Aid for Scientific
Research from the Ministry of Education, Culture, Sports, Science and
Technology, No.16340040.


\end{document}